# Enumeration and Online Library of Mass-Action Reaction Networks

Anastasia C. Deckard, Frank T. Bergmann, and Herbert M. Sauro*

Department of Bioengineering, University of Washington, Seattle, WA 98195-5502

ACD: cryptovolans@gmail.com

FTB: fbergman@u.washington.edu

HMS: hsauro@u.washington.edu (*Corresponding author)

Tel: USA 206-685-2119

Web site: www.sys-bio.org

# Abstract

## *Background*

The aim of this work is to make available to the community a large collection of mass-action reaction networks of a given size for further research. The set is limited to what can be computed on a modern multi-core desktop in reasonable time (< 20 days).

## *Results*

We have currently generated over 47 million unique reaction networks. All currently generated sets of networks are available and as new sets are completed they will also be made available. Also provided are programs for translating them into different formats, along with documentation and examples. Source code and binaries for all the programs are included. These can be downloaded from http://www.sys-bio.org/networkenumeration.

## *Conclusions*

This library of networks will allow for thorough studies of the reaction network space. Additionally, these methods serve as an example for future work on enumerating other types of biological networks, such as genetic regulatory networks and mass-action networks that include regulation.

# Background

In this paper we describe the enumeration of a significant number of reaction networks that follow mass-action kinetics. The aim of this work is to make available to the community a database of reaction networks for further research.

In previous work we developed methods for evolving reaction networks *in silico* with specific functional capabilities such as oscillators, bistable switches, and frequency filters [1, 2]. However, this work sampled only a very small proportion of all networks with a given functionality and we were intrigued by the possibility of enumerating the entire space of all networks. In this first paper we describe the approach we took to enumerate all reaction networks up to a given size. We begin by defining reactions networks, representing reaction networks using graph theory, followed by a detailed description of the algorithm used and a summary of the results. In the discussion we give one small application of the data involving the identification of network that can show oscillatory behavior.

## *Reaction Networks*

A reaction network is described by a set of reactions and associated molecular species. The simplest reaction networks are governed by mass-action kinetics, where the reaction rate of any particular reaction is proportional to the product of participating species, each raised to a corresponding kinetic order. Such reactions are also termed elementary reactions because they cannot be described in terms of simpler reactions. Given the elementary nature of the reactions, the kinetic orders will often correspond to the stoichiometries of the participating reactants, although in reality this need not be the case depending on the exact kinetic mechanism. In this work we assume that the kinetic orders correspond exactly to the stoichiometries. Given a generalized reaction such as:

$$n_1 A + n_2 B + \ldots \longrightarrow$$

where *A* and *B* are reactants with corresponding stoichiometries $n_1$ and $n_2$. We assume, without loss of generality, that all reactions are irreversible and that the rate law (as found through empirical observation) has the form:

$$v = kA^{n_1}B^{n_2}\ldots$$

where each reactant is raised to the power of the absolute value of its stoichiometric coefficient and *k* is the kinetic rate constant. For example, the forward reaction rate for the following reaction could be written as:

$$2ADP \rightarrow ATP + AMP$$

$$v = k\ ADP^2$$

A given reaction network will be made up of one or more reactions and molecular species. Figure 1 illustrates two typical reaction networks made up of a series of linked elementary reactions. Note that some species are connected by more than one reaction, often such species are termed internal species. Other species may only take part in a single reaction, these are often called boundary species. During a time course simulation of reaction networks any boundary species are often fixed in a simulation whereas the internal (sometimes called floating species) will evolve in time as the simulation progresses. In this paper we will not be concerned with simulation and therefore the distinction is less important.

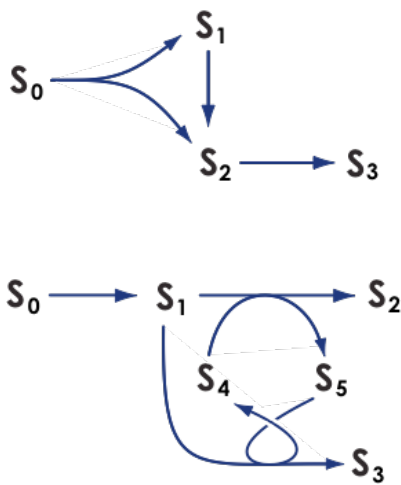

**Figure 1. Examples of reaction networks.**

### *Representing Reaction Networks as Graphs*

When working with reaction networks, it is convenient to represent reaction networks as graphs. A graph is made up of a collection of vertices (also called nodes or points), which represent entities, and edges (also called arcs or lines), which define the relationships between the entities. Graph theory defines several different types of graphs, the most important for our purposes being simple graphs or multigraphs, graphs with undirected or directed edges, and bipartite graphs. Simple graphs do not allow more than one edge between the same two vertices or for edges to form a loop on the same vertex, while multigraphs allow both of these to exist. Directed edges indicate an ordering between two connected vertices, usually drawn as arrows, while undirected edges do not indicate any ordering between two connected vertices. In a bipartite graph, there exist two distinct types of vertices and edges can only exist between vertices of different types. Many types of graphs have been used to represent and study chemical reaction networks [3]. For example, simple, undirected graphs have been used to study the relationships between metabolites, where the metabolites are the vertices and the edges indicate if two metabolites participate in the same reaction [4]. However, to represent a reaction network in a way that allows for studying its dynamics, it is necessary to capture certain key features. The edges must be directed to indicate the role of the species in the reaction, where a species vertex with an edge pointing towards a reaction vertex is a reactant and a species vertex with an edge coming from a reaction is a product. As a bipartite graph, the species vertices and reactions vertices are unique entities, and edges can only exist between one species vertex and one reaction vertex. All edges incident on one reaction vertex creates one reaction. The multigraph allows multiple edges between a given pair of reactions and species, which can therefore be used to represent autocatalysis. The reaction networks are therefore represented as directed bipartite multigraphs. This representation allows us to faithfully preserve the information about the structure of a chemical network by allowing us to represent not just simple unimolecular reactions (uni-uni) but also more complex reactions such as bimolecular reactions (uni-bi, bi-uni, bi-bi) as illustrated in Figure 2.

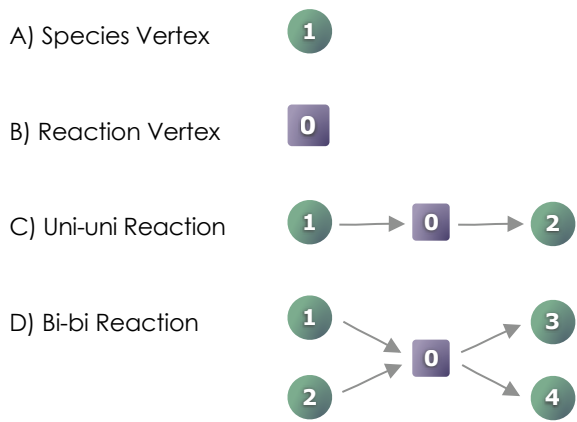

**Figure 2. Species and reactions as directed bipartite graphs.**

Species and reactions shown as vertexes (A and B) with distinct identities, as required in a bipartite graph. A uni-uni reaction (C) with two species. A bi-bi reaction (D) with four species.

In this work we will only consider reactions that contain up to two reactants and two products. In addition, all stoichiometries will be limited to one. However in reactions such as A+A →, which are permitted under these rules, the effective stoichiometry for species A is 2. The constraints on the types of allowable reactions means that reaction networks cannot include systems that contain reactions such as A+B+C → or A+A+B →. The reasons for these limits is that we consider tri-molecular reactions or higher to be extremely rare in real systems. For example, a reaction such as A+B+C → implies the simultaneous collision of three species which, probabilistically, is highly unlikely. Instead, such reactions are assumed to be non-elementary and can be constructed from our simpler set of allowed reactions. For example, the reaction A+B+C → can be represented using two more elementary reactions such as: A+B → AB; AB+C →. This formulation permits reaction networks of arbitrary complexity to be constructed. An example of a reaction network using the graph notation is shown in Figure 3.

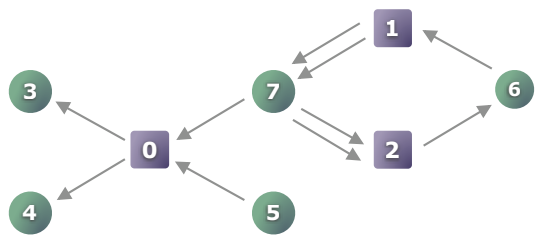

**Figure 3. Reaction network as a directed bipartite graph.**

A reaction network with three reactions and five molecular species. The reactions in this network are: 6 → 7 + 7, 7 + 7 → 6, and 5 + 7 → 3 + 4.

We use a concise format for storing the networks, which we will briefly explain here to aid researchers who wish to work with these networks. Reactions are represented as a list of pairs of vertices, where the first vertex is the source and the second vertex is the target. Each pair represents a directed edge between a reaction vertex and a species vertex. For example, the uni-uni reaction in Figure 2C can be represented by its edges: 1 → 0, 0 → 2, or more concisely as 1 0 0 2. The bi-bi reaction in Figure 2D could be represented as 1 → 0, 2 → 0, 0 → 3, 0 → 4, or more concisely as 1 0 2 0 0 3 0 4. For example, the network in Figure 3 has three reactions and five species, which represented as a graph would have three reaction vertices, labeled 0 to 2, and five species vertices, labeled 3 to 7. This network would be represented as 0 3 0 4 5 0 7 0 6 1 1 7 1 7 2 6 7 2 7 2. For all the networks stored in the repository, the number of reaction and the number of species are added to the beginning: **3 5** 0 3 0 4 5 0 7 0 6 1 1 7 1 7 2 6 7 2 7 2.

The structure of the network, or topology, is defined by the number and types of vertices, the number of edges, and how these vertices and edges are connected. Note that the characteristics associated with individual vertices (e.g. the concentration of a species or rate constant for a reaction) and individual edges are not part of the topology and are not considered when enumerating networks. Species and reactions in these networks do not have an identity, that is any species or reaction vertex could be swapped with another of the same kind and the topology would still be preserved. Thus this type of graph is called unlabeled. Additional care must be taken with unlabeled graphs when attempting to determine if the topology of the network is unique, or more technically, that the network is not isomorphic to another network. Isomorphisms exist when a given graph can be mapped to another graph in such a way that they are shown to have an identical structure when the identities of their vertices and edges are ignored [5]. More specifically, graphs *G* and *H* are isomorphic if there is a permutation *p* of vertex labels such that *p*(*G*) = *H*. As we are ignoring the identities of the components and we want to have only unique structures, we must therefore detect and ignore any isomorphic topologies.

### *Enumerating Reaction Networks*

Some work has been performed on generating reaction networks. For example, a computer application called Netscan [6] generates reaction networks from a specified list of reactions and that satisfy a required function. This application is capable of generating both metabolic and macromolecular networks. Another group has written software to enumerate reaction networks by building them from a set of simple and composite reactions, such as isomerization, dimerization, free radical combination, oxidation, autocatalysis, and so on [7]. The remainder of this paper explains our methods for enumerating all possible nonisomorphic graphs that represent reaction networks of a given size and the results obtained.

## Methods: Defining the Networks

To translate a reaction network into a directed bipartite multigraph, we must enforce constraints on the vertices and edges in the graph. We cover three groups of constraints: those that specify the type and size of networks to enumerate, those that ensure the reactions are valid, and finally those that ensure the reaction networks are correct and minimal.

### *Specifying the Network Set*

When attempting a large enumeration, it helps to break the problem into smaller subproblems. In the case of reactions networks, the most logical way to group networks is by the number of species, the number of reactions, and the types of reactions they possess. This also aids in making comparisons between the networks and identifying non-unique network structures. We will use the term "network set" to define the group of all possible, but unique, networks that have the specified number of species, number of reactions, and types of reactions. For example, all unique networks that have uni-uni, uni-bi, bi-uni, and/or bi-bi reactions, five species, and three reactions are considered a set. While the number of species and reactions are obvious, we will explain how the types of reactions are constrained by the max in|out degree in more detail.

*Max In|Out Degree* – This specifies the types of reactions that can exist in a network by restricting the maximum number of edges terminating on (called the *indegree*) or originating from (called the *outdegree*) a reaction vertex. For example, a max in|out degree of two means the maximum indegree = 2 and the maximum outdegree = 2. This would restrict the reactions to being bi-bi or less, which would allow bi-bi, uni-bi, bi-uni, and uni-uni reactions. While we only use a max in|out degree of two, it is possible to use larger values: a max in|out degree of three would restrict the reactions to being tri-tri or less, which would allow tri-tri, tri-bi, tri-uni, etc.

### *Constraints on the Reactions*

*Degree of Reaction Vertices* – The degree of the vertex is the number of edges connected to that vertex. In a directed graph, the degree is the sum of the indegree (the number of edges terminating on the vertex) and the outdegree (the number of edges originating from the vertex). Here, reaction vertices must have a degree of at least two: a minimum outdegree of one to connect the reaction to the product and a minimum indegree of one to connect the reaction to the reactant. The max in|out degree sets the maximum indegree and the maximum outdegree to the same value. For undirected graphs, the degree of reaction vertices must be between 2 * min in|out degree = 2 and 2 * max in|out degree. For directed graphs, the indegree and outdegree must be examined separately, where the indegree and outdegree must each be between min in|out degree = 1 and max in|out degree.

*No Self-Loops* – A symmetric self loop (e.g. A → A, A + A → A + A) could be removed without affecting the underlying chemical system and the network would have less reactions than the specified network set. Therefore, these types of reactions are omitted from our networks.

*No Hidden Species* – Certain reactions can be constructed such that they require hidden species to maintain mass conservation. If these hidden species are shown, then the structure of the reaction may be identical to another type of reaction. Therefore, we must exclude reactions with these hidden species where they would duplicate the structure of another type of reaction that we allow. We omit the reactions A + A → A and A → A + A because if their hidden species were shown they would become A + A → A + B and A + B → A + A. We also omit the reactions A + B → A and A → A + B because if their hidden species were shown they would become A + B → A + C.

*No Useless Reactions* – Reactions that are symmetric (e.g. A + B → A + B), could be removed without affecting the underlying chemical system and the network would have fewer reactions than the specified network set. Therefore, these types of reactions are omitted from our networks.

These constraints, when taken together, restrict the types of reactions that can be constructed. For example, given that four species exist and the max in|out degree is two, then a total of twelve different kinds of reactions can be enumerated (Figure 4).

**Figure 4. Possible reactions that can exist in enumerated reaction networks.**

The twelve different types of reactions (column A) and their directed graph representations (column B) that can be constructed given at least four species and a max in|out degree of two. The undirected graphs that underlie these reactions are also shown (columns C and D).

### *Constraints on the Network*

*No Redundant Reactions* – Redundant reactions are two or more identical reactions between the same species. If there are redundant reactions, then the duplicated reactions could be removed without affecting the underlying chemical system and the network would have less reactions than the specified network set. Therefore, networks that contain redundant reactions are not allowed.

*Connectivity* – Every graph must be connected, such that there exists a path between every pair of vertices. If it is disconnected, then it is two smaller networks and should not be considered part of the specified network set.

*Degree of Species Vertices* – There is no upper limit for the degree of the species vertices, but there must be a degree of at least one. This guarantees that each species is involved in at least one reaction, otherwise the network would violate the connectivity constraint.

## Methods: Enumerating the Networks

For a given network set (max in|out degree, number of species, and number of reactions), the following steps are performed to generate all possible networks: generate undirected bipartite simple graphs, transform these into undirected bipartite multigraphs, and finally transform these into directed bipartite multigraphs. Each of these steps uses the graphs generated in the previous step, therefore any constraints enforced in a previous step are perpetuated in subsequent steps.

For the first two steps, the software applications used are part of Nauty [8] by Brendan McKay, which is a collection of tools that use computationally efficient algorithms for isomorphism testing and graph generation. While there are several algorithms for graph isomorphism testing, such as the SD, Ullmann, and VF2 algorithms [9], Nauty tends to outperform these algorithms for large and/or densely-connected graphs with less regularity in their structure. This matches our expectations of reaction networks as being less regular in their connections and ensures that high performance is maintained as increasingly larger and denser networks are generated. Nauty, which performs the actual isomorphism testing, forms the backbone of a collection of related tools that are included with the Nauty package. These tools, called gtools, include programs for generating nonisomorphic graphs [10]. Nauty and the gtools, including their source code and user's guides, are available for download [11]. For the third and final step, a C++ program was developed to generate directed bipartite multigraphs and test for isomorphisms. To illustrate the process, we will use the network set with max in|out degree = 2, species = 2, and reactions = 2.

### *Generate Undirected Bipartite Simple Graphs*

The first step is to generate all possible undirected bipartite simple graphs that are nonisomorphic, which is performed by one of Nauty's gtools called genbg [10, 11]. Simple graphs are those where an edge between the same two vertices cannot be duplicated. To generate the graphs, each possible way to draw lines between the vertices is enumerated. The program enforces that the graph is connected and that each reaction vertex's degree is between the minimum and maximum. As in all steps, we require that all possible graphs that meet all the requirements are generated and are nonisomorphic. For example, specifying a max in|out degree = 2, species = 2, and reactions = 2 enforces that each reaction vertex's degree is between two and four. This yields only one undirected simple graph, shown in Figure 5A.

### *Transform into Undirected Bipartite Multigraphs*

In the second step, the simple graphs are used to generate all possible undirected bipartite multigraphs that are nonisomorphic, which is performed by one of Nauty's gtools called multig [10, 11]. Multigraphs are those where an edge between the same two vertices may be duplicated. The number of edges between the same two vertices is called the edge multiplicity. To generate the graphs, each simple graph is taken and all possible permutations of multiplicities of edges are created for each of its existing edges. The program enforces that for each edge the edge multiplicity is between one and the maximum degree and that each reaction vertex's degree is between the minimum and maximum. As in all steps, we require that all possible graphs that meet all the requirements are generated and are nonisomorphic. For example, specifying a max in|out degree = 2, species = 2, and reactions = 2 enforces that the edge multiplicity is between one and four and each reaction vertex's degree is between two and four. This yields thirteen undirected simple graphs, shown in Figure 5B.

Note that an edge with the maximum multiplicity would cause the reaction vertex to exceed the max in|out degree because no self-loops are allowed, so at least one other edge connects to a different species vertex. If self loops are allowed, all the edges for a single reaction vertex would connect to a single species vertex and the maximum multiplicity would be acceptable. However, we keep the multiplicity constraint because this allows for handling of self-loops if they were included and the reaction vertex degree constraint will properly handle either case.

### *Transform into Directed Bipartite Multigraphs*

In the final step, the undirected multigraphs are used to generate all possible directed bipartite multigraphs that are nonisomorphic, which is performed by a C++ program. To generate the graphs, each undirected multigraph is taken and the first edge is directed one way and then the other, for both of these the second edge is directed one way and then the other, etc. until all permutations of the directed edges are generated. Therefore there are $2^{\#undirected\ edges}$ possible networks that could be generated for each topology, but we reduce this number by verifying the networks as they are generated. If the program detects multiple edges between the same two vertices, then the order of the directed edges does not matter and the program only generates combinations instead of permutations for the directed edges. For example, when the program directs two edges in opposite directions between the same two vertices it will only generate the combination →← instead of generating the permutations →← and ←→, as these two permutations have the same meaning. A network is removed if: 1) The indegree or outdegree of a reaction vertex exceeds the max in|out degree. 2) The indegree or outdegree of a reaction vertex is less than the min in|out degree. 3) The network contains any useless reactions, such as A + B → A + B. 4) The network contains any redundant reactions, such as A → B and A → B.

If the network has met all these constraints, it is considered a valid chemical network and is then checked against other valid networks that were generated from the same undirected bipartite multigraph to determine if it is nonisomorphic. The most straightforward way to find an isomorphism between two networks is to look at all the permutations of vertex labels in the first network to see if the permutation creates a set of edges that are the same as the second network's set of edges [12]. While this graph isomorphism testing is exponential in nature and therefore a computationally intensive method, several basic properties work to significantly reduce the work that needs to be performed. For the vertex label permutations, the list of permutations is generated once for the given set of networks as they all share the same number of species and vertices. The graphs are bipartite, so only vertices of the same type can be interchanged (reaction vertex labels are only swapped with other reaction vertex labels and species vertex labels are only swapped with other species vertex labels). Compared to a non-bipartite graph, this reduces the number of permutations from (reactions + species)! to reactions!*species!. For isomorphism testing, it is possible to avoid more costly comparisons by looking at the degrees of the vertices. If two networks do not have the same number of vertices with the same values for the degrees then the networks cannot be isomorphic and we can move on to checking the next network. Conversely, if the vertices do have matching degrees, then each label permutation of the first network must be compared to the second network. With the label permutation that maps the vertices of the first network to the vertices of the second network, if the corresponding vertices do not have the same degree then the label permutation will not generate an isomorphism, so comparing the edges can be skipped and the algorithm can move on to checking the next label permutation. These additions have greatly increased the speed of the software. For example, checking the degrees of the label permutation made the software over 30 times faster, with greater speed gains occurring in the larger sets we tested.

Once the program starts generating valid networks, it maintains a list of valid nonisomorphic networks. The first valid network that is generated is put into this list, and each subsequent valid network that is generated is compared to each nonisomorphic network that is already in the list. Once all the networks have been generated and tested, they are saved to a file. The program then reads in the next undirected multigraph and repeats this process until all the undirected multigraphs have been transformed into directed multigraphs. The final file contains all the directed multigraphs that represent all possible biochemical networks with the defined number of species and reactions. For example, specifying a max in|out degree = 2, species = 2, and reactions = 2 yields the 36 networks shown in Figure 5C.

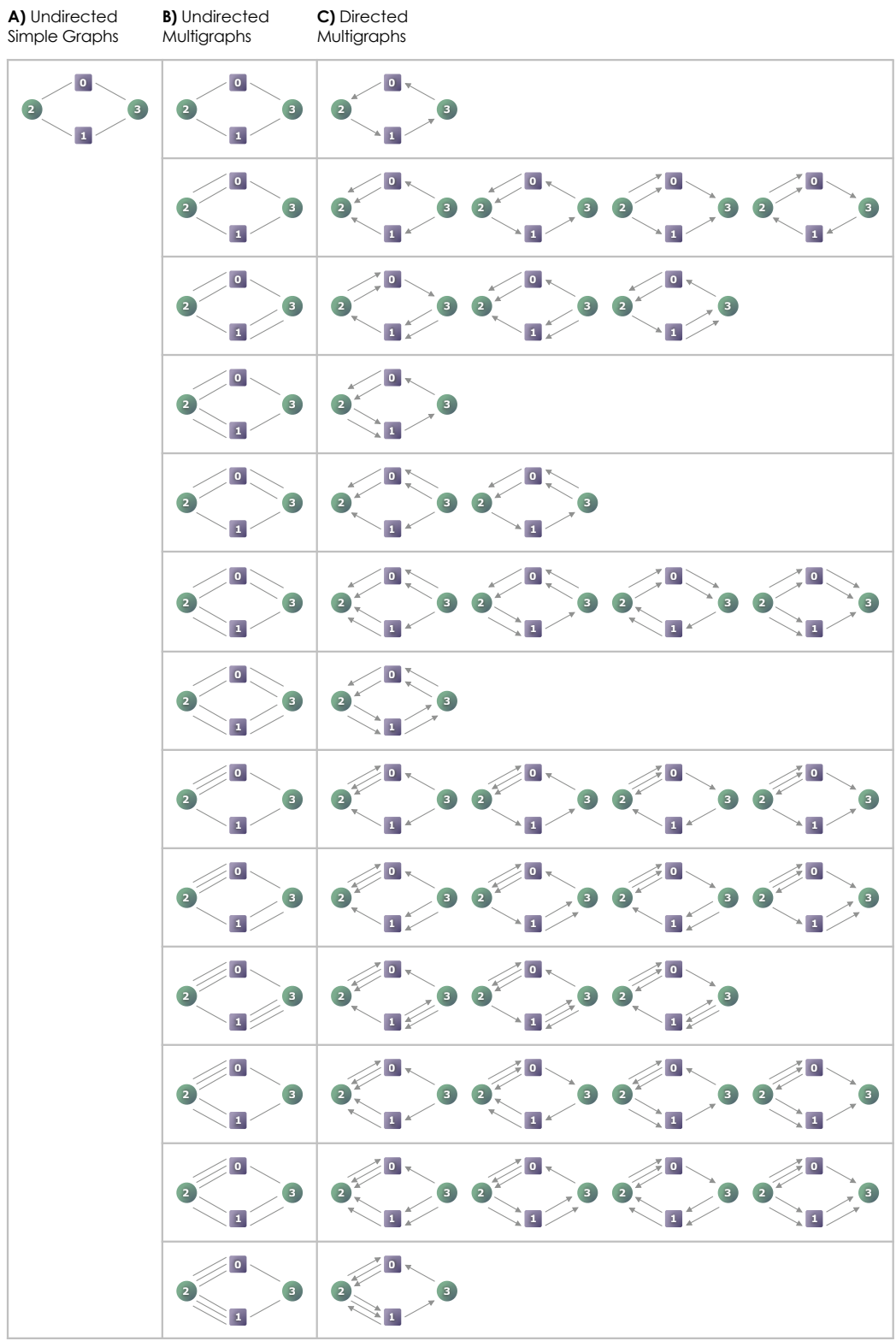

**Figure 5. Enumerated reaction networks for two species and two reactions.**

Column C shows all possible directed bipartite multigraphs for max in|out degree = 2, species = 2, and reactions = 2 that meet the constraints, which are the final chemical networks. Columns A and B show the graphs from which they were derived.

## Methods: Optimizing Network Enumeration

As discussed previously, the problem size and runtime grows exponentially with the number of edges in the bipartite multigraphs. Thus optimization of the program was a crucial feature to push the boundaries of what could be enumerated with currently available desktop hardware.

We began by optimizing iterator usage in the program, as well as porting it to the Linux platform (specifically Ubuntu 8.10). In the next step we examined ways to use mulitcore processors available in currently available hardware. OpenMP, the Open Multi Processing API for multi-platform multiprocessing, seemed the best choice for this [13, 14]. The enumeration problem itself is highly amenable to parallelization as the set of directed multigraphs generated for one undirected multigraph is disjoint from any set generated for another undirected multigraph. The major cost factor in the enumeration is the validation that a network is indeed nonisomorphic to all other networks. As indicated above, in the worst case reactions!*species! permutations have to be evaluated. This large number of permutations can be partitioned into parallel batches and the results compared in stages. Exploiting these possibilities with OpenMP made the program approximately two times faster on dual processor machines and four times faster on quad processor machines.

Currently we are evaluating CUDA, the Compute Unified Device Architecture, a SDK developed by NVIDIA [15]. CUDA offers the possibility of running hundreds of threads in parallel on the GPU rather than the CPU, which promises a chance for computing yet another round of species and reactions. However, a first prototype proved disappointing, as the amount of time spend copying data from the host machine to the graphics card proved prohibitive.

For completeness sake, it should be mentioned that MPI, the Message Passing Interface API, and the use of clusters was not employed. The sole reason being that the communication overhead for the isomorphism testing would have been prohibitively large. Alternatively, the individual nodes of the cluster would have to be equipped with far more memory to hold all permutations.

## Results

We have generated all networks with two to ten species and one to three reactions, two to nine species with four reactions, two to four species with five reactions, and two species with seven reactions, for a total of 47,471,353 networks (Table 1). While we were initially concerned by the large numbers of networks we would be working with, we were pleased to find that our optimizations were able to greatly speed up the most computationally-expensive portion of the program: the isomorphism testing. Even though the label permutations (reactions!*species!) increase steeply as the size of the networks increases, the increase in the number of isomorphism tests as the network size increases was kept manageable (Table 2).

|  |  | Reactions | | | | | | |
|---|---|---|---|---|---|---|---|---|
|  |  | 1 | 2 | 3 | 4 | 5 | 6 | 7 |
| S p e c i e s | 2 | 6 | 36 | 110 | 255 | 396 | 472 | 396 |
|  | 3 | 5 | 218 | 4082 | 52679 | 527339 | 4306169 |  |
|  | 4 | 1 | 274 | 19335 | 796303 | 24159763 |  |  |
|  | 5 | 0 | 136 | 30246 | 3276608 |  |  |  |
|  | 6 | 0 | 29 | 21913 | 5651388 |  |  |  |
|  | 7 | 0 | 3 | 8526 | 5068577 |  |  |  |
|  | 8 | 0 | 0 | 1865 | 2666292 |  |  |  |
|  | 9 | 0 | 0 | 225 | 877692 |  |  |  |
|  | 10 | 0 | 0 | 14 |  |  |  |  |

**Table 1. Numbers of enumerated reaction networks.**

| Species | Undirected Simple Graphs | Undirected Multigraphs | Possible Directed Graphs | Valid Networks | Label Permutations (s! * r!) | Isomorphism Tests | Isomorphic Networks | Final Networks |
|---|---|---|---|---|---|---|---|---|
| 2 | 1 | 32 | 49856 | 160 | 12 | 136 | 50 | 110 |
| 3 | 6 | 336 | 572416 | 5184 | 36 | 10502 | 1102 | 4082 |
| 4 | 22 | 879 | 1692160 | 24692 | 144 | 88129 | 5357 | 19335 |
| 5 | 38 | 931 | 2045568 | 43640 | 720 | 274337 | 13394 | 30246 |
| 6 | 45 | 584 | 1455104 | 40860 | 4320 | 658419 | 18947 | 21913 |
| 7 | 33 | 231 | 650240 | 23228 | 30240 | 856318 | 14702 | 8526 |
| 8 | 18 | 65 | 206848 | 8712 | 241920 | 1360074 | 6847 | 1865 |
| 9 | 6 | 12 | 43008 | 2160 | 2177280 | 1669047 | 1935 | 225 |
| 10 | 2 | 2 | 8192 | 432 | 21772800 | 1432690 | 418 | 14 |

**Table 2. A comparison of the numbers that are involved in enumerating reaction networks with three reactions.**

The first column shows the number of species for the networks, all of which have three reactions. The next two columns show the numbers of undirected simple graphs and multigraphs that are generated. Possible Directed Graphs shows how many directed multigraphs could have been generated, while the Valid Network column shows how many directed multigraphs were actually generated (which represent valid chemical networks). These networks are then subject to isomorphism testing. Label Permutations show the number of different ways the vertices could be relabeled, i.e. how different ways an isomorphism could exist. Isomorphism Tests shows the actual number of tests where comparing edges between networks had to be performed, but this number is reduced by preliminary comparisons of degrees. Isomorphic Networks were those that were found and removed. Final Networks shows the final count of reaction networks with the given number of species and reactions.

All materials can be found at http://www.sys-bio.org/NetworkEnumeration. For the data, the sets of generated networks and associated data are provided and newer sets will be made available as they complete. For the software, we are providing source code and binaries for Windows, Unix, and Mac OS X. All Software is licensed under the BSD license. The software consists of the programs for enumerating all possible networks for a given number of species and reactions (dependent on a modified version of Brendan McKay's nauty suite and with optional OpenMP support) and programs for translating encoded networks into the model exchange languages SBML [16] and Jarnac Script [17] and directly into stoichiometry matrices. This software was written in standard C and C++. Finally we provide online services, in the form of a web application and web service, that translates encoded networks into different formats (The online applications were written in C# and rely on the Systems Biology Workbench running on the server back-end). For documentation, a description of the storage format for the networks, information on the enumeration algorithm, and guides for the programs are also included.

While our methods have been successful for enumerating reaction networks so far, we are rapidly approaching the limit of what can be enumerated within reasonable time and memory requirements. We plan on improving the isomorphism testing with more advanced algorithms, in the hope that this will allow us to the generate a few more of the larger sets before we reach the inevitable limit of what can be computed given then current state of desktop hardware.

## Discussion

The next step in exploring the search space is to develop methods and software that will examine each network to determine if it is capable of the desired functionality (oscillator, homeostatic network, bistable, etc.). A given program will be able to read the specified network format, simulate the network, evolve the rate constants for the network with the specified objective function, and output if the network met its objective, and if so, store the parameters.

As a preliminary test of this method, we have used a modified version of one of our previous evolutionary algorithms [2] to test networks for oscillatory behavior. We tested all networks with four species and three reactions, for a total of 19,335 networks, and found four networks with oscillatory behavior (Figure 6). It is known that the smallest mass-action network that can exhibit a Hopf bifurcation is composed of three floating species and five reactions [18]. Therefore we did not expect any oscillators of this kind in this group of networks. However Lotka-Volterra oscillators, i.e conservative oscillators, can be generated from just two variables. In the Lotka-Volterra oscillator, the initial conditions will affect the amplitude of the oscillations (unlike an oscillator generated from a Hopf bifurcation) even though no conserved cycles exist in the network. The four networks shown in Figure 6 are of the Lotka-Volterra type with one predator species and one prey species [19]. The two floating (internal) species are the prey and predator, and the two fixed (boundary) species are constant and can therefore be folded into the rate constants. While our preliminary testing of the networks was not exceptionally rigorous, we consider it a proof of concept for our continued work in this area.

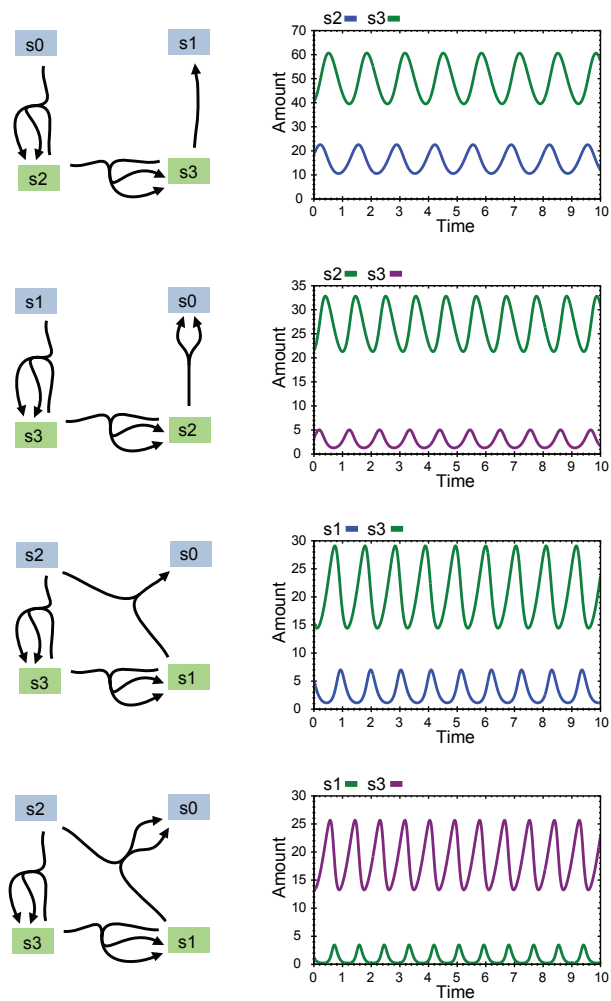

**Figure 6. Oscillators with four species and three reactions.**

Oscillators found in the preliminary testing of the networks with four species and three reactions. All these models are essentially identical at the core. Peripheral changes, particularly reaction changes between the last two species (on the right in the networks), do not affect the dynamics. Interestingly there is one other variant missing which the evolutionary algorithm appears to have not identified. The computation to identify these networks took approximately three days working through approximately 20,000 networks on a desktop computer.

The largest obstacle to perform this sort of exhaustive testing of possible reaction networks is the huge amounts of time that are required. Even if we could determine whether a network is capable of a given function at a rate of one network per second, it would take just over 15 years to check 500 million networks on one computer. Unfortunately, we do not even currently approach the rate of a network per second; Optimizing a network with our evolutionary algorithm to determine oscillations required approximately 30 seconds per network, which means it would take approximately over 475 years to check 500 million networks on one computer. It is therefore necessary to employ faster algorithms that can examine the structural properties of the networks and deduce their dynamic characteristics. These algorithms should be able to identify whether a network is capable or incapable of a specific functionality. For example, algorithms exist for examining the structure of the network to determine if the network does or does not have the capacity for multiple equilibria [20]. Using this type of algorithm would allow us to exclude networks from further testing, which would avoid the costly act of simulating and/or optimizing the network.

# Conclusions

The step of enumerating all possible unique mass-action reaction networks provides the data for use in future studies. With over 47 million networks for several different sizes of networks, and more to added as they are enumerated, this collection of networks will allow for thorough studies of the reaction network space. Additionally, these methods serve as an example for future work on enumerating other types of biological networks, such as genetic regulatory networks and mass-action networks that include regulation.

# Authors' Contributions

ACD defined the graphs to represent the desired reactions networks, modified/created programs for enumerating graphs, and drafted manuscript. FTB optimized the enumeration program, added support for parallel processing, and helped to draft the manuscript. HMS defined the types of reaction networks to enumerate, verified that the graphs represented reaction networks, and helped to draft the manuscript. ACD and HMS conceived of the study and participated in its design and coordination. All authors have read and approved the final manuscript.

# Acknowledgments

The programs used to generate the nonisomorphic graphs, Nauty, genbg, and multig, were created by Brendan McKay (http://cs.anu.edu.au/~bdm/nauty/) and we would like to thank him for his help with his programs. The authors would like to thank Michal Galdzicki, Sean Sleight, and Lucian Smith for running the enumeration software on their computers. We would also like to thank the NSF 0527023- *FIBR* for their generous funding.

# References

1. Deckard A, Sauro HM: **Preliminary studies on the in silico evolution of biochemical networks**. *Chembiochem* 2004, 5(10):1423–31.

2. Paladugu SR, Chickarmane V, Deckard A, Frumkin JP, McCormack M, Sauro HM: **In silico evolution of functional modules in biochemical networks**. *IEE Proceedings-Systems Biology* 2006, 153(4):223–35.

3. Temkin ON, Zeigarnik AV, Bonchev D: *Chemical Reaction Networks: A Graph-Theoretical Approach*. Boca Raton: CRC-Press; 1996:41–59.

4. Wagner A, Fell DA: **The small world inside large metabolic networks**. *Proc. R. Soc. London B* 2001, 268(1478): 1803-1810.

5. Ettinger M: **The complexity of comparing reaction systems**. *Bioinformatics* 2002, 18 (3):465–469.

6. Clarke B, Fawcett G, Mittenthal JE: **Netscan: a procedure for generating reaction networks by size**. *Journal of Theoretical Biology* 2004, 230:591-602.

7. Ramakrishnan N, Bhalla US: **Memory switches in chemical reaction space**. *PLoS Comput Biol*. 2008, 4(7):e1000122.

8. McKay B: **Practical Graph Isomorphism**. *Congressus Numerantium* 1981, 30:45–87.

9. De Santo M, Foggia P, Sansone C, Vento M: **A large database of graphs and its use for benchmarking graph isomorphism algorithms**. *Pattern Recognition Letters* 2003, 24:1067–1079.

10. McKay B: **Isomorph-Free Exhaustive Generation**. *Journal of Algorithms* 1998, 26(2): 306–324.

11. **Nauty** [http://cs.anu.edu.au/~bdm/nauty/]

12. Gross JL, Yellen J: *Graph Theory and Its Applications*. Boca Raton: CRC Press; 2006:89–97.

13. Dagum L, Menon R: **OpenMP: An Industry-Standard API for Shared-Memory Programming**. *IEEE Computational Science & Engineering* 1998, 5(1):46–55

14. Gabriel E, Fagg GE, Bosilca G, Angskun T, Dongarra JJ, Squyres JM, Sahay V, Kambadur P, Barrett B, Lumsdaine A, Castain RH, Daniel DJ, Graham RL, Woodall TS: **Open MPI: Goals, Concept, and Design of a Next Generation MPI Implementation**. In *Recent Advances in Parallel Virtual Machine and Message Passing Interface*. Berlin: Springer; 2004:97–104

15. **NVIDIA's CUDA** [http://www.nvidia.com/cuda]

16. Hucka M, Finney A, Sauro HM, Bolouri H, Doyle JC, Kitano H, Arkin AP, Bornstein BJ, Bray D, Cornish-Bowden A, Cuellar AA, Dronov S, Gilles ED, Ginkel M, Gor V, Goryanin II, Hedley WJ, Hodgman TC, Hofmeyr JH, Hunter PJ, Juty NS, Kasberger JL, Kremling A, Kummer U, Le Novere N, Loew LM, Lucio D, Mendes P, Minch E, Mjolsness ED, Nakayama Y, Nelson MR, Nielsen PF, Sakurada T, Schaff JC, Shapiro BE, Shimizu TS, Spence HD, Stelling J, Takahashi K, Tomita M, Wagner J, Wang J: **The systems biology markup language (SBML): a medium for representation and exchange of biochemical network models**. *Bioinformatics* 2003, 19(4):524–531.

17. Sauro HM: **JARNAC: a system for interactive metabolic analysis**. In *Animating the Cellular Map*. Edited by: Hofmeyr J-HS, Rohwer JM, Snoep JL. Stellenbosch, University Press Stellenbosch; 2001:221-228.

18. Wilhelm T, Heinrich R: **Smallest chemical reaction system with Hopf bifurcation**. *Journal of Mathematical Chemistry* 1995, 17:1–14.

19. Epstein IR, Pojman JA: *An Introduction to Nonlinear Chemical Dynamics: Oscillations, Waves, Patterns, and Chaos*. New York: Oxford University Press; 1998:4–5.

20. Craciun G, Feinberg M: **Multiple equilibria in complex chemical reaction networks: I. The Injectivity Property**. *SIAM Journal on Applied Mathematics* 2005, 65(5):1526–1546.